\documentclass{ws-procs975x65}

\usepackage{graphicx,epsfig, amsmath,amssymb,amsfonts}

\begin{document}


\title{SUPERNOVAE CONSTRAINTS ON DGP MODEL AND COSMIC TOPOLOGY\footnote{This 
research has been partially supported by CNPq.}} 

\author{MARCELO J. REBOU\c{C}AS} 
\address{Centro Brasileiro de Pesquisas F\'{\i}sicas\\
Rua Dr.\ Xavier Sigaud 150, \  22290-180 Rio de Janeiro --
RJ, Brazil}

\begin{abstract}
We study the constraints that the detection of a non-trivial spatial 
topology may place on the parameters of braneworld models by considering 
the Dvali-Gabadadze-Porrati (DGP) and the globally homogeneous Poincar\'e 
dodecahedral spatial (PDS) topology as a circles-in-the-sky observable 
topology. To this end we reanalyze the type Ia supernovae constraints on 
the parameters of the DGP model and show that PDS topology gives rise to 
strong and complementary  constraints on the parameters of the DGP model. 
\end{abstract}

\bodymatter

\section{Introduction}\label{intro}

In the  standard cosmology, the Universe is
described by a space-time manifold $\mathcal{M}_4 = \mathbb{R}
\times M_3$ endowed with a locally (spatially) homogeneous and 
isotropic  metric
\begin{equation}
\label{RWmetric}
ds^2 = - dt^2 + a^2 (t) \left [ \frac{dr^2}{1-kr^2} + r^2 (d\theta^2
+ \sin^2 \theta \, d\phi^2) \right ] \,,
\end{equation}
where, depending on the spatial curvature $k$,
the geometry of the $3$--space $M_3$ is either Euclidean ($k=0$),
spherical ($k=1$), or hyperbolic ($k=-1$).
The spatial section $M_3$ is usually taken to be one of the 
simply-connected spaces: Euclidean $\mathbb{R}^3$, spherical $\mathbb{S}^3$, 
or hyperbolic $\mathbb{H}^3$. 
However, given that the connectedness of the spatial sections
$M_3$ has not been determined by cosmological observations, and
since geometry does not fix the topology, our $3$--dimensional space
may be one of the possible multiply connected quotient
manifolds  of the form $\mathbb{R}^3/\Gamma$, $\mathbb{S}^3/\Gamma$, and
$\mathbb{H}^3/\Gamma$, where $\Gamma$ is a fixed-point free group of
isometries of the corresponding covering space.
Thus, for example, for the Euclidean geometry ($k=0$) besides $\mathbb{R}^{3}$
there are 6 classes of topologically distinct compact orientable spaces $M_3$. 

The immediate observational consequence of a detectable nontrivial 
topology\cite{TopDetec} of $M_3$ is the existence
of the circles-in-the-sky,\cite{CSS1998} i.e., pairs of 
matching circles will be imprinted on the CMBR anisotropy sky maps%
~\cite{CSS1998}. Hence, to observationally probe a putative
nontrivial topology of $M_3$, one should examine the full-sky CMBR 
maps in order to extract the pairs of correlated circles and determine
the spatial topology.

In the context of the 5D braneworld models the universe is described 
by a $5$-dimensional metrical orbifold (bulk) $\mathcal{O}_5$ that is 
mirror symmetric ($\mathbb{Z}_2$) across the 4D  brane (manifold) 
$\mathcal{M}_4$. Thus, the bulk can be decomposed as  $\mathcal{O}_5 
= \mathcal{M}_4 \times E_1 =  \mathbb{R} \times M_3 \times \mathbb{E}_1$,
where $E_1$ is a $\mathbb{Z}_2$ symmetric Euclidean space, and  where 
$\mathcal{M}_4$ is endowed with a Robertson--Walker metric~(\ref{RWmetric}), 
which is recovered when $w=0$ for the extra non-compact spatial dimension. 
In this way, the multiplicity of possible inequivalent topologies of 
our $3$--dimensional space, and the physical consequences of a
non-trivial detectable topology of $M_3$ as the circles-in-the-sky 
are brought on the braneworld scenario.

Here we briefly study the constraints that a detection of a spatial 
topology may place on the parameters of a simple braneworld 
modified-gravity model that accounts for the accelerated expansion of 
the universe via infrared modifications to general relativity, namely the Dvali-Gabadadze-Porrati (DGP) model\cite{DGPmodel}, as generalized to 
cosmology by Deffayet\cite{Deffayet2000}.  To this end we reanalyze the 
type Ia supernovae constraints on the parameters of the DGP model and 
show that PDS topology gives rise to strong and complementary  constraints 
on the parameters of the DGP model.

\section{Constraints and Concluding Remarks} \label{Constr-Concl} 

{}Using the first-year data the WMAP team reported a total density
value\cite{WMAP-Spergel:2003} $\Omega_{\mathrm{tot}}=1.02 \pm\, 0.02$, 
while the three-year WMAP article\cite{WMAP-Spergel:2006} reports six 
different values for the $\Omega_{\mathrm{tot}}$ ranging from a very 
nearly flat $\Omega_{\mathrm{tot}}= 1.003^{+0.017}_{-0.013}$ to positively 
curved $\Omega_{\mathrm{tot}}= 1.037^{+0.015}_{-0.021}$ depending on the
combination of data set used to resolve the geometrical degeneracy.

The Poincar\'e dodecahedral space (PDS), $\mathcal{D}=\mathbb{S}^3/I^\star$,
explains both the suppression of power of the low multipoles  and this 
observed  total density. We note, however, that other topologies as 
$\mathcal{O}=\mathbb{S}^3/O^\star$ also remain viable.\cite{Aurich12}
Attempts to find antipodal or nearly-antipodal circles-in-the-sky in the
WMAP data have failed~\cite{Cornish}. There is, however, claim of hints 
of matching circles\cite{Roukema} in ILC WMAP maps, which a second group 
has confirmed\cite{Key-et-al:2006} but have also shown that the circle 
detection lies below the false positive threshold.\cite{Key-et-al:2006}. 
On the other hand, even if one embraces 
the result that pairs of antipodal (or nearly antipodal) circles of
radius $\gamma \geq 5^\circ$ are undetectable in the current CMBR maps,%
\cite{Key-et-al:2006} the question arises as to whether the circles are 
not there or are merely hidden by various sources of contamination
(Doppler, integrated Sachs-Wolfe, e.g.), or even due to the angular 
resolution of the current CMBR maps, as suggested in 
Ref.~\refcite{Holger2006}. The answer to these questions requires 
great care, among other things, because the level 
of contamination depends on both the choice of the cosmological models 
(parameters) and on the topology\cite{Weeks2006}. Results so far remain
non-conclusive, i.e., one group finds their negative outcome to be robust 
for globally homogeneous topologies, including the dodecahedral space, in 
spite of contamination\cite{Key-et-al:2006}, while another group 
finds the contamination strong enough to hide the possible 
correlated circles in the current CMBR maps.\cite{Aurich2006,Holger2006}
%
\begin{figure}[!htb]
\begin{center} 
\includegraphics[width=4.5cm,height=4.5cm,angle=0]{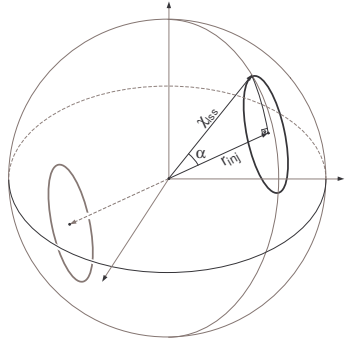}
\caption{\label{CinTheSky1} A schematic illustration of two antipodal
matching circles in the LSS. } 
\end{center}
\end{figure}

In  $\mathcal{D}$ space the pairs of matching circles are necessarily 
antipodal as shown in Fig.~\ref{CinTheSky1}.
Clearly the distance between the centers of each pair of the 
correlated  circles is twice the injectivity radius of the smallest sphere 
inscribable  $\mathcal{D}$.
A straightforward use of trigonometric relations for the right-angled 
spherical triangle shown in Fig.~\ref{CinTheSky1} yields 
\begin{equation}
\label{ChiLSS}
\chi^{}_{lss}= \frac{d^{}_{lss}}{a_0} = \sqrt{|\Omega_k|}
\int_1^{1+z_{lss}} \, \frac{H_0}{H(x)} \,\, dx \, =\, \tan^{-1} 
\left[\,\frac{\tan r_{inj}}{\cos \alpha}\, \right] \;,
\end{equation}
where $d^{}_{lss}$ is the radius of the LSS, $x=1+z$ is an 
integration variable, $H$ is the Hubble 
parameter, $\Omega_k =1-\Omega_{\mathrm{tot}}$, $r_{inj}$ is a 
topological invariant (equals to $\pi/10$ for $\mathcal{D}$), 
the distance $\chi^{}_{lss}$ is measured \emph{in units of the
curvature radius}, 
$a_0=a(t_0)=(\,H_0\sqrt{|1-\Omega_{\mathrm{tot}}|}\,)^{-1}\,$,
and $z_{lss}=1089$~\cite{WMAP-Spergel:2003}.

Equation~(\ref{ChiLSS}) makes apparent that $\chi^{}_{lss}$ depends on
the cosmological scenario. For the DGP model one has
\begin{equation} \label{DGP}
\left({{H} \over {H_0}}\right)^2=
\Omega_k(1+z)^2+\left(\sqrt{\Omega_m(1+z)^3+\Omega_{r_c}}+\sqrt{\Omega_{r_c}}\right)^2,
\end{equation}
where $r_c$ is a length scale beyond which gravity starts to leak out 
into the bulk.
Equations~(\ref{ChiLSS}) and~(\ref{DGP}) give the relation between 
the angular radius $\alpha$ and the parameters of the DGP model, and thus 
can be used to set constraints on these parameters. 

To illustrate the role of the cosmic topology in constraining the DGP
parameter we consider the $\mathcal{D}$ spatial topology, and assume  
the angular radius $\alpha = 50^\circ$ and uncertainty 
$\delta {\alpha} \simeq 6^\circ$.  
\begin{figure*}[ht!]
\begin{center}
\includegraphics[height=4.9cm]{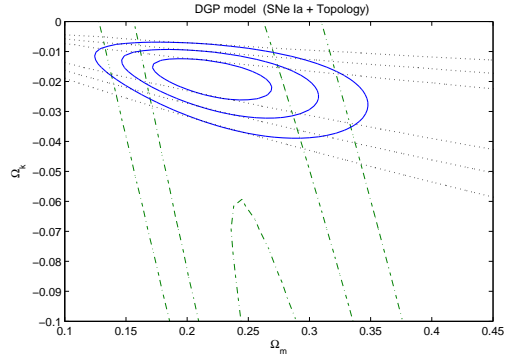} 
\caption{\label{DGP_p0}Confidence contours ($68.3\%$, $95.4\%$
and $99.7\%$) in the $\Omega_m-\Omega_k$ plane for DGP model obtained 
with the SNe Ia gold sample assuming a ${\cal D}$ space topology with 
$\gamma=50^o\pm6^\circ$. Also shown are
the contours obtained assuming no topological data (dash-dotted
lines) and the ones corresponding to topology only (dotted lines).}
\end{center}
\end{figure*} 
Figure~\ref{DGP_p0} shows the results of our joint SNe Ia plus cosmic
topology analysis, where the {\em gold} sample of 157 SNe Ia, as compiled 
by Riess {\em et al.},\cite{Riess2004} was used.
There we display the confidence regions in the 
parametric plane $\Omega_{k}\,$--$\,\,\Omega_{m}$ and also the 
regions from the conventional analysis with no such a topology
assumption.
The comparison between these regions makes clear that
the effect of the $\mathcal{D}$ topology is to reduce considerably 
the area corresponding to the confidence intervals in the parametric 
plane as well as to break degeneracies arising from the current SNe 
Ia measurements. The best-fit parameters for this joint analysis are
$\Omega_m = 0.232$ and $\Omega_k =-0.018\,$.

For a detailed analysis of topological constraints in the context of 
the DGP and other models, including braneworld inspired models, see Refs.~\refcite{Previous1}  and Refs.~\refcite{Previous2}.
Finally, we note that in Refs.~\refcite{Fairbairn-et-al} constraints
are placed on the DGP models using supernova and other data but 
with no such a topological constraint.

\section*{Acknowledgments} 
Valuable discussions with M.C. Bento, O. Bertolami and N.M.C. Santos
are gratefully acknowledged. I am grateful to A.F.F. Teixeira 
for indicating misprints and omissions.

\end{document}